\documentclass[aps,prd,a4paper,floatfix,tighten,nofootinbib,preprint,showpacs]{revtex4-1}
\usepackage{graphicx}
\usepackage{epic}
\usepackage{eepic}
\usepackage{latexsym}
\usepackage{upgreek}

\newcommand{\be}{\begin{equation}}
\newcommand{\ee}{\end{equation}}
\newcommand{\bea}{\begin{eqnarray}}
\newcommand{\eea}{\end{eqnarray}}

\newcommand{\ga}{\gamma}
\newcommand{\pp}{\phantom}

\def\al{\alpha}
\def\b{\beta}

\newcommand{\de}{\delta}

\newcommand{\si}{\tilde{\sigma}}

\def\na{\nabla}

\begin{document}

\title{Stress-Energy Tensor of the Quantized Massive Fields in Friedman-Robertson-Walker Spacetimes.}

\author{Jerzy Matyjasek}
\email[]{jurek@kft.umcs.lublin.pl}
\affiliation{Institute of Physics, Maria Curie-Sk{\l}odowska University\\
pl. Marii Curie-Sk\l odowskiej 1, 20-301 Lublin, Poland}
\author{Pawe{\l} Sadurski}
\affiliation{Institute of Physics, Maria Curie-Sk{\l}odowska University\\
pl. Marii Curie-Sk\l odowskiej 1, 20-301 Lublin, Poland}

\date{\today}

\begin{abstract}
The approximate stress-energy tensor of the quantized massive scalar, 
spinor and vector fields in the spatially flat Friedman-Robertson-Walker 
universe is constructed. It is shown that for the scalar fields with 
arbitrary curvature coupling, $\xi,$ the stress-energy tensor calculated 
within the framework of the Schwinger-DeWitt approach is identical 
to the analogous tensor constructed in the adiabatic vacuum. Similarly, 
the Schwinger-DeWitt stress-energy tensor for the fields of 
spin 1/2 and 1 coincides with the analogous result calculated by 
the Zeldovich-Starobinsky method. The stress-energy tensor thus obtained
 are  subsequently used in the back reaction problem. It is 
shown that for pure semiclassical Einstein field equations with 
the vanishing cosmological constant and the source term consisting 
exclusively of its quantum part there are no self-consistent exponential 
solutions driven by the spinor and vector fields. A similar situation takes 
place for the scalar field if the coupling constant belongs 
to the interval $\xi \gtrsim 0.1.$  For a positive cosmological 
constant the expansion slows down for all considered types of massive fields
except for minimally coupled scalar field.
The perturbative approach to the problem is briefly discussed and
possible generalizations of the stress-energy tensor are indicated.

\end{abstract}


\maketitle

\section{Introduction}
In their recent paper~\cite{Kaya},  Kaya and Tarman constructed the stress-energy tensor,
$T_{ab},$ of the massive  quantized scalar field with the arbitrary curvature coupling, $\xi,$ in  
the spatially flat ($k=0$) Friedman-Robertson-Walker spacetime, described by the line element
\begin{equation}
 ds^{2} = a^{2}(\eta)\left(-d\eta^{2} + dw^{2}+dy^{2}+dz^{2} \right)
\label{frwl}
\end{equation}
within the framework of the adiabatic regularization~\cite{Parker1,Parker2,Parker3,Bunch1,Bunch2,Anderson,Landete}. 
Using sixth-order WKB approximation to the 
solutions of the covariant Klein-Gordon equation  
\begin{equation}
 - \Box \phi^{(0)} + (m^{2} +\xi R) \phi^{(0)} =0
\label{covKG}
\end{equation}
in the formally divergent expression for the $T_{ab},$ after  adiabatic regularization, 
they calculated the leading term of the  approximation to the components of the  (covariantly) 
conserved stress-energy tensor. In the adiabatic regularization one has to subtract, mode by 
mode, the infinite terms of the adiabatic order 0, 2 and 4. To avoid ambiguities it is necessary 
to subtract all terms that contain at least one divergent part for arbitrary parameters of the 
theory~\cite{ParTom}. 

Although the intermediate stages of the calculation are rather involved, the final 
result is surprisingly simple and can be schematically written in the form
\begin{equation}
 \frac{1}{96\pi^{2} m^{2} a^{12}}\left(\sum A_{ijk}^{pqs} \left[a^{(i)}\right]^{p}\left[a^{(j)}\right]^{q}\left[a^{(k)}\right]^{s} 
 + B a^{3}\dot{a}\ddot{a} a^{(3)}\right),
 \label{schemat}
\end{equation}
where $A$ and $B$ are numerical coefficients (possibly dependent on the coupling constant), 
the overdots denote differentiation with respect to the coordinate $\eta$ and
\begin{equation}
 a^{(k)}=\frac{d^{k} a(\eta)}{d\eta^{k}}  \hspace{0.5cm} {\rm with} \hspace{0.5cm}  a^{(0)} =a(\eta).
\end{equation}
The summation is extended over all $i,...$ and $p,...$ satisfying
$ pi+qj+ks =6,$ Additionally, the number of appearances of the function 
$a(\eta)$ and its derivatives in each term is 6, i.e., $p+q+s=6.$ The same 
is true for the last term in Eq.~(\ref{schemat}).

Because of the spatial symmetries it suffices to calculate the energy density, $\rho = -T_{0}^{0},$
as the remaining components are easily calculated from the equations
\begin{equation}
 T_{x}^{x} \equiv T_{1}^{1} = T_{2}^{2} = T_{3}^{3}
 \label{symm}
\end{equation}
and
\begin{equation}
 T_{x}^{x} = T_{0}^{0}+ \frac{a}{3 \dot{a}} \, \dot{T}_{0}^{0}.
 \label{conse}
\end{equation}
Alternatively, one can calculate the trace of the tensor, or, due to simplicity of the 
spatial mode functions for the $k=0$ geometries, the component $T_{x}^{x}.$ Of course 
the first method is simplest and the remaining ones may serve as a useful check of the 
calculations. 

It should be noted that being local, the adiabatic approximation does not give
much information about the particle creation. On  the other hand, it is very useful 
when the vacuum polarization effects dominate. In the adiabatic expansion  (for $k=0$) 
the number of derivatives plays the role of the perturbation parameter. It is expected 
that this procedure gives reasonable results provided 
$\dot{a}/a, \ddot{a}/a,... \ll \Omega =(m^2 a^2+\tilde{k}^{2} )^{1/2},$
where $0\leq \tilde{k} <\infty$~\cite{Grib}.

In this paper we shall demonstrate that the adiabatic tensor of Ref.~\cite{Kaya} can 
be obtained as the special case of the more general expression calculated within the framework 
of the Schwinger-DeWitt approach~\cite{DeWitt}.
The generalization can be twofold: one can allow for nonvanishing $k$ in the line element
and  consider the massive fields of spin 1/2 and 1. In what follows we shall restrict ourselves 
to the simplest case of the spatially flat Friedman-Robertson-Walker geometries, thus avoiding 
the subtleties connected with the Euclideanization of the line element. Like the adiabatic 
expansion, the  Schwinger-DeWitt method is local and ignores nonlocal phenomena such 
as particle creation, and, consequently, its domain of applicability is limited. 
Nevertheless, it gives reasonable results if the Compton length, $\lambda_{C},$ associated 
with the quantized field is much smaller than the characteristic radius of curvature. 

The adiabatic calculations are based on the WKB approximation to the mode functions
and their derivatives and  subsequent integration (summation) of the functions thus constructed. 
On the other hand, the Schwinger-deWitt approach may be considered as purely 
geometrical. Moreover, it can be demonstrated that for the massive spinor and vector
fields the method of Zeldovich and Starobinsky~\cite{Yakov,Grib} and Schwinger-DeWitt 
give the same results. (All  fields considered in this paper are neutral).
The equality of the results obtained from both methods is impressive. 
It should be noted that such equality must not be taken for granted: 
Indeed, it is expected that discussed methods give (approximate) Green functions with 
the same  structure of singularities as $x' \to x.$  However, this does not mean
the functions are the same.

The paper is organized as follows. In Section II, after giving a brief sketch of the 
Schwinger-DeWitt method, we construct the renormalized stress-energy tensor of the 
quantized massive fields in a large mass limit in the  spatially-flat Friedman-Robertson-Walker 
universe. The result (in a scalar case) is subsequently compared with the analogous tensor
obtained within the framework of the adiabatic method. We also briefly report on our 
calculations of the stress-energy tensor of the quantized massive s=1/2 and s=1 fields 
using the Zeldovich-Starobinsky method. This section contains the core results 
of the paper. In Section  III, to illustrate some possible applications of the  constructed 
tensors, two families of solutions to the semiclassical Einstein field equations  
(both exact and approximate) are constructed. Using the Routh-Hurwitz criterion we perform 
stability analysis. It is pointed out that the perturbative method adapted in this paper 
is, in this context, equivalent to the reduction of order used by  Parker and Simon~\cite{Simon}
Finally, in Sec. IV, among other things, 
we indicate possible generalizations of the results presented in this paper. Throughout the paper 
we use the MTW conventions~\cite{MTW}.

\section{The Schwinger-DeWitt Approach} 

The one-loop approximation to the effective action of the quantized massive fields 
in a large mass limit can be  constructed within the framework of the Schwinger-DeWitt 
method. The effective action of the scalar field described by  Eq.(\ref{covKG}) can be 
obtained form the  coincidence limit of the Hadamard-DeWitt coefficient $a_{3}.$ 
Unfortunately,  neither the covariant Dirac equation
\begin{equation}
(\ga^{a} \na_{a}\,+\, m) \phi^{(1/2)}\,=\,0,
\label{Dirac}
\end{equation}
nor the equation describing the massive vector field
\begin{equation}
 (\de^{a}_{b}\Box\,-\,\na_{b}\na^{a}\,-
 \,R^{a}_{b} \,-\,\de^{a}_{b}m^{2})\phi^{(1)}_{a}\,=\,0,
 \label{Vect}
\end{equation}
have the form required by the Schwinger-DeWitt method. 
In the first case one can introduce a new spinor, $\psi^{(1/2)}$ defined as 
$\phi^{(1/2)} =(\gamma^{a} \nabla_{a} -m) \psi^{(1/2)}.$ This, after commuting covariant 
derivatives  and making use of the elementary properties  of the $\gamma$ matrices, 
results in the second-order equation:
\begin{equation}
 \left(\nabla_{a}\nabla^{a}  -\frac{1}{4} R - m^{2}\right)\psi^{(1/2)}=0.
\end{equation}
To  eliminate the nondiagonal differential operator $\nabla_{b}\nabla^{a}$ in Eq.~(\ref{Vect}) 
one can employ the method of Barvinsky and Vilkovisky~\cite{Barvinsky,Frolov} and demonstrate 
that the effective action of the massive vector field equals the effective action calculated 
for Eq.~(\ref{Vect}) without the nondiagonal term minus the effective action of the minimally 
coupled scalar field. The approximate effective action of the quantized spinor, scalar and 
vector fields can be written as~\cite{Avramidi,IvanBook}
 \begin{eqnarray}
 W^{(1)}_{ren}\,&=& \,{1\over 192 \pi^{2} m^{2}} \int d^{4}x g^{1/2}
 \left( \al^{(s)}_{1} R \Box R
+\al^{(s)}_{2} R_{a b} \Box R^{a b}
+ \al^{(s)}_{3} R^{3}
+\al^{(s)}_{4} R R_{a b} R^{a b} 
\right. \nonumber \\
&&
+\al^{(s)}_{5} R R_{a b c d} R^{a b c d}        
+\al^{(s)}_{6} R^{\pp{b} a}_{b} R^{\pp{c} b}_{c} R^{\pp{a} c}_{a}
+\al^{(s)}_{7} R_{a b} R^{c d} R_{c \pp{a} d}^{\pp{c} a \pp{d}  b}
+\al^{(s)}_{8} R_{a b} R^{a}_{\pp{a} e c d} R^{b e c d}
\nonumber \\
&&\left. 
+\al^{(s)}_{9} R_{c d}^{\pp{c} \pp{d} a b} R_{a b}^{\pp{a} \pp{b} e h}
R_{ e h}^{\pp{e} \pp{h} c d}
+\al^{(s)}_{10} R_{c ~ d}^{~ a ~ b} R_{a ~ b}^{~ e ~ s} R_{e ~ s}^{~ c ~ d}
 \right)\nonumber \\
 &&={1\over 192 \pi^{2} m^{2}} \int d^{4}x g^{1/2} \sum_{i}^{10} \al_{i} {\rm Inv}_{i},
\label{dzialanie}
\end{eqnarray}
where $m$ is the mass of the field and the numerical coefficients depending on 
the spin of the field are  given in a Table~\ref{table1}. 
\begin{table}[h]
\begin{tabular}{|c|c|c|c|}\hline
& s = 0 & s = 1/2 & s = 1 \\\hline\hline
$\al^{(s)}_{1} $ & $ {1\over2}\xi^{2}\,-\,{1\over 5} \xi $\,+\,${1\over
56}$ & $- {3\over 280}$ &
$- {27\over 280}$\\
 $\al^{(s)}_{2}$ & ${1\over 140}$ & ${1\over 28}$ & ${9 \over 28}$\\
 $\al^{(s)}_{3}$ &$ \left( {1\over 6} - \xi\right)^{3}$& ${1\over 864}$ &$
- {5\over 72}$\\
 $\al^{(s)}_{4}$ & $- {1\over 30}\left( {1\over 6} - \xi\right) $& $-
{1\over 180}$ & ${31\over 60}$\\

 $\al^{(s)}_{5}$ & ${1\over 30}\left( {1\over 6} - \xi\right)$ &$ -
{7\over 1440}$ &$ - {1\over 10}$\\

 $\al^{(s)}_{6}$ & $ - {8\over 945}  $& $- {25 \over 756}$ & $- {52\over
63}$\\

 $\al^{(s)}_{7}$ & ${2 \over 315}$ & $ {47\over 1260}$  & $- {19\over 105} $\\
 $\al^{(s)}_{8}$ & ${1\over 1260}$ & ${19\over 1260} $ & ${61\over  140} $\\
 $\al^{(s)}_{9}$ & ${17\over 7560}$& ${29\over 7560}$ & $- {67\over 2520}$\\
 $\al^{(s)}_{10}$ & $- {1\over 270}$ & $ - {1\over 108} $  & $ {1\over 18}$\\ \hline
 \end{tabular}
 \caption{The coefficients $\al_{i}^{(s)}$ for the massive scalar with arbitrary
curvature coupling $\xi$ , spinor,
and vector
 field}
 \label{table1}
 \end{table}

The (renormalized) stress-energy tensor can be calculated from the standard relation
\begin{equation}
T^{ab} = \frac{2}{\sqrt{g}} \frac{\delta  W_{ren}^{(1)}}{\delta g_{ab}}
\label{set}
\end{equation}
and consists of the purely geometric terms constructed from the Riemann tensor, its covariant 
derivatives and contractions. The type of the field enters through the coefficients $\alpha_{i}.$
Additionally, each term
\begin{equation}
 \frac{2}{\sqrt{g}} \frac{\delta }{\delta g_{ab}} \int d^{4}x g^{1/2}  {\rm Inv}_{i}
\end{equation}
is covariantly conserved. In a general D-dimensional manifold the invariants ${\rm Inv}_{i}$ are 
independent and this is the reason why we prefer to work with this form of the action~\cite{ja2006}.
It should be noted, however, that the Riemann tensor satisfies numerous dimension-dependent
identities and some invariants in a specific dimension, say, $D=4,$ are not independent.
We shall briefly discuss this problem at the end of this section. It should be stressed
that regardless of the particular representation of the effective action the stress-energy
tensor for a given background and field should give unique results.

In Refs.~\cite{mat1,mat2} the most general form of the stress-energy tensor has been calculated. 
It consists of almost 100 terms of the background dimensionality 6 and, for obvious reason,  
it will be not presented here. In spite of the limitations mentioned above it is still the most 
general result. Moreover, by construction, it depends functionally on the metric tensor allowing, 
in principle, to solve the semiclassical Einstein field equations in a self-consistent way.

Defining $\tilde{\alpha}_{i} = \alpha_{i}/192 \pi^{2} m^{2}$ in (\ref{dzialanie}) and treating 
them as free parameters one arrives at the action functional considered by Lu and Wise~\cite{Lu} 
in the black hole context\footnote{ Actually, Lu and Wise consider the action functional with 
$\tilde{a}_{1} = \tilde{a}_{2} =0$ and a different organization of indices in the last term term,
which, however, is equal to the last term of (\ref{dzialanie}) plus 1/4 the $\tilde{a}_{9}$ term.} 
On the other hand,  by retaining only $\alpha_{9}$ term in (\ref{dzialanie}) one obtains the
Goroff-Sagnotti effective action~\cite{Goroff}  studied by Dobato and Maroto~\cite{Dobado}.

For the line element (\ref{frwl}) the ``time'' component of the stress-energy tensor has the form 
\begin{eqnarray}
 &&T_{0}^{0}=\frac{1}{32 \pi^{2} m^{2} a^{12}}\left[(22 \al_{10} - 72 \al_{2} - 84 \al_{4} - 168 \al_{5} - 85 \al_{6} - 13 \al_{7} - 70 \al_{8} - 84 \al_{9}) \dot{a}^{6}\right.\nonumber \\
 &&+ 12 (36 \al_{1} - 3 \al_{10} + 22 \al_{2} + 12 \al_{4} + 24 \al_{5} + 12 \al_{6} + 2 \al_{7} + 10 \al_{8} + 12 \al_{9}) a \dot{a}^{4}\ddot{a}\nonumber \\
 &&-2 (156 \al_{1} - 3 \al_{10} + 62 \al_{2} + 12 \al_{4} + 24 \al_{5} + 12 \al_{6} + 2 \al_{7} +  10 \al_{8} + 12 \al_{9}) a^{2} \dot{a}^{3} a^{(3)}\nonumber \\
&&-(306 \al_{1} - 3 \al_{10} + 142 \al_{2} + 540 \al_{3} + 192 \al_{4} + 204 \al_{5} + 87 \al_{6} + 47 \al_{7} + 70 \al_{8} + 72 \al_{9}) a^{2} \dot{a}^{2} \ddot{a}^{2}
\nonumber \\
&& -2(18 \al_{1} + 8 \al_{2} + 36 \al_{3} + 12 \al_{4} + 12 \al_{5} + 5 \al_{6} + 3 \al_{7} + 4 \al_{8} + 
 4 \al_{9}) a^{3} {a^{(2)}}^{3} \nonumber \\
 &&-4 (3 \al_{1} + \al_{2}) a^{4} \dot{a} a^{(5)}+2 (6 \al_{1} + 2 \al_{2}) a^{4}\ddot{a} a^{(4)} -2 (3 \al_{1}  + \al_{2}) {a^{4} a^{(3)}}^{2} +32 (3 \al_{1} + \al_{2}) a^{3} \dot{a}^{2} a^{(4)}\nonumber \\
 &&+\left. 6 (18 \al_{1} + 8 \al_{2} + 36 \al_{3} + 12 \al_{4} + 12 \al_{5} + 5 \al_{6} + 3 \al_{7} + 4 \al_{8} + 
 4 \al_{9}) a^{3} \dot{a} \ddot{a} a^{(3)} \right],
 \label{comp_tt}
 \end{eqnarray}
where, for typographical reasons, the superscripts describing spin of the field have been omitted.
It can be shown that substituting the $\alpha^{(0)}$ parameters into the $T_{0}^{0}$ component of the 
stress-energy tensor one gets precisely the result obtained by Kaya and Tarman. Further, making use 
of (\ref{conse}) and (\ref{comp_tt}) one obtains the spatial components which are identical to those 
obtained within the framework of the adiabatic expansion.~\footnote{We have recalculated stress-energy 
tensor of the massive scalar field using the adiabatic approach. There are two typographical errors 
in the expression for spatial components of the stress-tensor as presented in~\cite{Kaya}. With these 
terms corrected, the adiabatic tensor is identical to the tensor constructed within the framework of the 
Schwinger-DeWitt method.} Although the remaining components of the stress-energy tensor can be calculated 
using the methods described above we shall  display them for readers' convenience:
\begin{eqnarray}
&&T_{x}^{x} =\frac{1}{96 \pi^{2} m^{2} a^{12}}\left[- 9 \left(22 \alpha_{10}-72 \alpha_{2}-84 \alpha_{4}-168 \alpha_{5}-85 \alpha_{6}-13 \alpha_{7}-70 \alpha_{8}-84 \alpha_{9} \right) \dot{a}^{6} \right.
\nonumber \\
&&- 6 \left(576 \alpha_{1}-70 \alpha_{10}+424 \alpha_{2}+276 \alpha_{4}+552 \alpha_{5}+277 \alpha_{6}+45 \alpha_{7}+230 \alpha_{8}+276 \alpha_{9}\right) a \dot{a}^4 \ddot{a}
\nonumber \\
&&+ 2 \left(1308 \alpha_{1}-39 \alpha_{10}+566 \alpha_{2}+156 \alpha_{4}+312 \alpha_{5}+156 \alpha_{6}+26 \alpha_{7}+130 \alpha_{8}+156\alpha_{9}\right) a^{2} \dot{a}^{3} a^{(3)}
\nonumber \\
&&- 2 \left(444 \alpha_{1}-3 \alpha_{10}+158 \alpha_{2}+12 \alpha_{4}+24 \alpha_{5}+12 \alpha_{6}+2 \alpha_{7}+10 \alpha_{8}+12 \alpha_{9}\right) a^{3} \dot{a}^{2} a^{(4)}
\nonumber \\
&&+ 2 \left(120 \alpha_{1}+46 \alpha_{2}+108 \alpha_{3}+36 \alpha_{4}+36 \alpha_{5}+15 \alpha_{6}+9 \alpha_{7}+12 \alpha_{8}+12 \alpha_{9}\right) a^{4} \ddot{a} a^{(4)}
\nonumber \\
&&-2 \left(198 \alpha_{1}-3 \alpha_{10}+94 \alpha_{2}+324 \alpha_{3}+120 \alpha_{4}+132 \alpha_{5}+57 \alpha_{6}+29 \alpha_{7}+46 \alpha_{8}+48 \alpha_{9}\right) a^{3} \ddot{a}^{3}
\nonumber \\
&&+ 2 \left(69 \alpha_{1}+29 \alpha_{2}+108 \alpha_{3}+36 \alpha_{4}+36 \alpha_{5}+15 \alpha_{6}+9 \alpha_{7}+12 \alpha_{8}+12 \alpha_{9}\right) a^{4} {a^{(3)}}^{2}
\nonumber \\
&&+ 5 \left(774 \alpha_{1}-33 \alpha_{10}+410 \alpha_{2}+756 \alpha_{3}+384 \alpha_{4}+516 \alpha_{5}+237 \alpha_{6}+85 \alpha_{7}+194 \alpha_{8}
\right. \nonumber  \\
&& + \left. 216 \alpha_{9}\right) a^{2} \dot{a}^{2} \ddot{a}^{2}
- 2 \left(1098 \alpha_{1}-12 \alpha_{10}+472 \alpha_{2}+1188 \alpha_{3}+444 \alpha_{4}+492 \alpha_{5}+213 \alpha_{6}
\right. \nonumber \\
&&+107 \left. \left.\alpha_{7}+172 \alpha_{8}+180 \alpha_{9}\right) a^{3} \dot{a} \ddot{a} a^{(3)}    
+  52 \left(3 \alpha_{1}+\alpha_{2}\right) a^{4} \dot{a} a^{(5)} - 4 \left(3 \alpha_{1}+\alpha_{2}\right) a^{5} a^{(6)} \right].
\label{comp_xx}
\end{eqnarray}
Now, by substituting $\alpha_{i}^{(1/2)}$ and $\alpha_{i}^{(1)}$ into Eqs.~(\ref{comp_tt}) and (\ref{comp_xx})
one obtains the renormalized stress-energy tensor of the spinor and vector fields, respectively.
To the best of our knowledge the results for  the  spinor and vector fields reported here are new.

Having established that the two different approaches give the same result for the scalar field, the natural 
question is whether it is also true for the fields of spin 1/2 and 1. In order to answer this question 
we have used asymptotic expansion of the Green function constructed within the framework
of the Zeldovich-Starobinsky approach~\cite{Yakov,Beilin,Grib}. Performing integrations over momenta
and summations over spin states we obtained, after much algebra,  the results that are identical 
to those calculated within the framework of the Schwinger-DeWitt method. The Mathematica notebooks
with the details of this calculation for $s=0, 1/2$ and $1$ fields are available upon request
from the first author.

It is of some interest, especially when one wants to address the back reaction problem 
perturbatively, to restore the physical constants. In that case  the multiplicative factor 
standing in front of the integral~(\ref{dzialanie}) is $\hbar \lambda_{C}^{2}/192 \pi^{2}$ 
and the integrand has the dimension $ {\rm Length}^{-6}.$ In order keep control of the order 
of terms in complicated expansions we shall introduce, whenever useful, a small parameter
$\varepsilon$ which should be set to 1 at the end of the calculations.

We conclude this section with some remarks about the approximate action (\ref{dzialanie}). In four 
dimensions there are two additional identities~\cite{Xu1,Harvey}, which can be directly obtained from 
\begin{equation}
 R^{ab}_{\phantom{a}\phantom{b}[ab} R^{cd}_{\phantom{c}\phantom{d}cd} R^{e}_{\phantom{e}e]} =0
\end{equation}
and
\begin{equation}
 R^{ab}_{\phantom{a}\phantom{b}[ab} R^{cd}_{\phantom{c}\phantom{d}cd} R^{ei}_{\phantom{e}\phantom{i}ei]} =0.
\end{equation}
It means that the 8th and 10th term in the effective action can be expressed as linear combinations of 
the remaining ones. Consequently, the same result can be obtained simply by substituting  
$
\sum_{i=1} ^{n=10} \al_{i} {\rm Inv}_{i} 
$
in Eq.~(\ref{dzialanie}) by 
$
\sum_{i=1} ^{n=10} \tilde{\b}_{i} {\rm Inv}_{i}, 
$
where 
$\tilde{\b}_{1} =\al_{1},$ $\tilde{\b}_{2} = \al_{2},$
$\tilde{\b}_{3} =\al_{3} +1/4 \al_{8} -5/8 \al_{10},$
$\tilde{\b}_{4} =\al_{4} -2 \al_{8} +9/2 \al_{10},$
$\tilde{\b}_{5} =\al_{5} +1/4 \al_{8} -3/8 \al_{10},$
$\tilde{\b}_{6} = \al_{6} +2 \al_{8} -4 \al_{10},$
$\tilde{\b}_{7} =\al_{7} + 2 \al_{8} -3 \al_{10},$
$\tilde{\b}_{8} =0,$ 
$\tilde{\b}_{9} = \al_{9} +1/2 \al_{10}$ and $\tilde{\b}_{10}=0.$
Moreover, one can further simplify the effective action of the quantized fields
in a Weyl-flat ($C_{abcd} =0$) spacetime making use of the invariants
constructed from the Weyl tensor. It is possible provided the functional derivative
of such invariant with respect to the metric tensor vanishes.
For example functionals constructed from  
$C_{abcd} C^{cd}_{\phantom{c}\phantom{d}ij} C^{ijab},$ $R C_{abcd}C^{abcd}$
and $C_{abcd}\Box C^{abcd}$ have the desired property. 

\section{The Back Reaction}
In the purest form of the semiclassical approximation, the gravitational field 
is treated classically, but it is driven by the mean value of $T_{a}^{b}.$ 
Having constructed the leading term of the renormalized stress-energy tensor 
which depends functionally on a generic metric, one can analyze the semiclassical Einstein 
field equations  with the total stress-energy tensor consisting of the classical 
and quantum parts. It should be emphasized once more that accepting the 
approximation (\ref{dzialanie}) we ignore particle creation, which is a nonlocal 
effect and concentrate on the vacuum polarization effects. 

Further, to simplify our discussion we shall assume that the renormalized 
coupling parameters $k_{1}$ and $k_{2}$ in the quadratic part of the total 
action
\begin{equation}
S_{q} =\int d^{4} x \sqrt{-g} \left( k_{1} R_{ab}R^{ab} + k_{2}  R^{2}\right),
\label{kwa}
\end{equation}
identically vanish. In the case on hand the tensors obtained by functional 
differentiation of the integrated $R^{2}$ and $R_{ab}R^{ab}$ with respect to
the metric tensor are not independent but  there is an additional term
\begin{equation}
 k_{3}\left( -\frac{1}{12} R^{2} g_{ab} +R^{cd}R_{cadb}\right)
\label{kwb}
 \end{equation}
which can be added  to the left had side of the equations. In the latter we put $k_{3} =0.$
The semiclassical Einstein field equations have, therefore, a standard form
\begin{equation}
G_{ab}[g] +\Lambda g_{ab} =8\pi  T^{(total)}_{ab}[g], 
                      \label{semiE}
\end{equation}
where  $T^{(total)}_{ab} = T_{ab}^{(class)} + T_{ab},$ 
i.e., the right hand side of (\ref{semiE}) is a sum
of the classical and quantum stress-energy tensor. 

Let us postpone the further discussion of the semiclassical equations 
for a while and return to the quadratic terms. It should be noted that
such terms appear in a natural way as a low-energy limit of the string
action. Indeed, they appear as the first-order correction to the 
classical action in $\alpha'$ expansion ($\alpha' = 2 \pi \lambda_{S}^{2}$
where $\lambda_{S}$ is a fundamental length). In a general parametrization
in D-dimensions the action should be supplemented by a Kretschmann scalar,
$R_{abcd}R^{abcd}.$ Functionally differentiating the total action with
respect to the metric tensor and taking into account the $(0,0)$ component
of the thus obtained tensor equation, which usually is the simplest one 
to analyze, it can be demonstrated~\cite{shapiro1} that for a constant dilaton field 
there is no self-consistent exponential solution in $D=2$ and $D=4.$ 
The latter result, that is of immediate relevance here,  is a simple consequence of 
the fact that the quadratic terms vanish for the spatially-flat Friedman-Robertson-Walker 
spacetime with the exponential scale factor whereas the Einstein tensor does not.

If the total stress-energy tensor is known, one can, in principle, construct 
the self-consistent solution of the system~(\ref{semiE}). It should be noted however, 
that since  the general stress-energy tensor, $T_{a}^{b},$ is constructed from the coincidence 
limit of $a_{3}(x,x'),$ it contains the terms with higher-order derivatives of 
$g_{ab},$ and, consequently, there is a real danger that the semiclassical 
equations may lead to physically unacceptable solutions. Moreover, 
the tensor $T_{a}^{b}$  and the resulting equations are rather complicated 
and it is natural that one is forced to look for simplifications. 
First, let us consider the function $a(t)$ which also satisfies the additional 
relation~\cite{Dobado}
\begin{equation}
 a'(t) = c_{1}^{1/2} a(t),
\label{row}
 \label{reduction}
\end{equation}
where $t$ is ``ordinary'' time coordinate related to $\eta$ by $dt = a(\eta) d\eta,$
$c_{1}^{1/2}$ is some constant and a prime denotes differentiation with 
respect to $t.$ 
Transforming the semiclassical equations (\ref{semiE}) 
to $(t,w,y,z)$ coordinates and making use of (\ref{reduction}) one obtains
\begin{equation}
 -3 c_{1} + 3 \tau c_{1}^{3} + \Lambda =-8 \pi \rho
 \label{se}
\end{equation}
and the similar equation with the right hand side substituted by $8 \pi p,$
where
\begin{equation}
 \tau=\frac{1}{12 \pi m^{2}}\left( 144 \al_{3}+36\al_{4} +24 \al_{5} +9\al_{6} +9 \al_{7} 
 + 6 \al_{8}+4 \al_{9} + 2 \al_{10}\right).
\label{tau}
\end{equation}
From the semiclassical equations one has either $p =\sigma \rho$ with $\sigma=-1$ and $\rho \ne 0$ 
(which is the equation of state for the cosmological constant) or $\rho=p=0.$ Since $\sigma =-1$ 
the only nondegenerate equation of state compatible with the simplified semiclassical equations one can 
consider an effective cosmological constant $\Lambda_{eff}= \Lambda+ 8\pi \rho.$

In the absence of both $\Lambda$ and $\rho$ (or $\Lambda_{eff}$) one has either $c_{1} =0$ 
or $c_{1}^{2} =1/\tau.$
The right hand side of the above equation  is negative for spinors and vectors, and, 
consequently, there are no exponential solutions driven by massive spinors and 
vectors in this simple model.  The scalar case is slightly more complicated because 
of the coupling constant $\xi.$  Indeed, inspection of (\ref{tau}) shows that it is 
positive for $\xi < \xi_{crit} =0.1023$ and hence there is no solution for the 
conformally coupled scalar field. The temporal evolution of the model is governed 
by (\ref{row}). For $c_{1}^{1/2} >0$ one has 
\begin{equation}
 a(t) = a(t_{0}) \exp(c_{1}^{1/2} (t-t_{0})),
\end{equation}
whereas for $c_{1} =0$ one has static universe with the constant scale factor.
A few words of comment are in order. First observe 
that the quantum part of the tensor can be made arbitrarily large 
simply by taking large  number of fields. Indeed, for  $N$ fields 
of a given spin, $s$ with masses $m_{i}$ the  renormalized effective 
action is still of the form (\ref{dzialanie}) with
\begin{equation}
 \frac{1}{m^{2}} \to \sum_{i=1}^{N} \frac{1}{m_{i}^{2}}.
\end{equation}
Moreover, even if the $c_{1}^{2}$ is negative for some  fields  it does not mean that the field
should be excluded from the model. It can contribute to the total stress-energy tensor,
which yields a  proper overall characteristics.

Let us return to the semiclassical equations (\ref{se}) and (\ref{tau}) with $\Lambda_{eff} \neq 0.$ The third order equation 
has exact solutions, but they are not particularly illuminating. It is of some interest to 
consider them in some important regimes. Since the present value of the cosmological constant 
is extremely small one has
\begin{equation}
 H_{0}^{2} \equiv c_{1} \approx \frac{\Lambda_{eff}}{3}\left(1 + \frac{\Lambda_{eff}^{2}\tau}{9} \right),
 \label{at1}
\end{equation}
\begin{equation}
  H_{0}^{2} \equiv c_{1} \approx  \frac{1}{\sqrt{\tau}} - \frac{\Lambda_{eff}}{6}
  \label{at2}
\end{equation}
and
\begin{equation}
  H_{0}^{2} \equiv c_{1} \approx - \frac{1}{\sqrt{\tau}} - \frac{\Lambda_{eff}}{6}.
\end{equation}
Following~\cite{tomislav} we shall call (\ref{at1}) the classical de Sitter (or to be more precise 
the quantum-corrected classical) solution whereas (\ref{at2}) is the quantum solution.
Now, a natural question arises about the stability of (\ref{at1}) and (\ref{at2}) against small 
perturbations. To answer this question let us express the full set of the semiclassical 
equations (\ref{semiE}) in terms of $H(t) = a'(t)/a(t),$ insert  $H(t) = H_{0} + \delta H(t)$ 
(where $\delta H(t)$ is small perturbation) and finally linearize the thus constructed
equations. Assuming $\delta H(t) = \exp(\lambda t)$ one obtains the characteristic 
polynomial(s) $F(\lambda)$ with the real coefficients. The solution is stable under 
small perturbations if all roots of the characteristic polynomial lie in the left 
complex half-plane. A useful criterion has been given by Routh and Hurwitz: 
If the determinants of all upper left-corner minors of the Hurwitz matrix are
strictly positive then all roots of the polynomial $F(\lambda)$ have negative 
real parts. Sometimes it is more efficient to use the necessary condition  which 
requires all the coefficients of $F(\lambda)$ to be positive~\cite{cesari}.
Of course, if the order of the characteristic polynomial is low one can solve
the equation exactly and check if the conditions $Re(\lambda_{i})<0$ are satisfied. 

Let us return to our case. It is advantageous not to specify the values of $H_{0}$ 
and analyze the relations among the characteristic polynomials $F_{t}(\lambda),$
$F_{x}(\lambda)$ and $F(\lambda),$ where $F_{t}$ is obtained from the $(0,0)$
component of the semiclassical equations, $F_{x}$ from the spatial component and
$F =F_{x} + \sigma F_{t},$ i.e.,  is linear  combination of the former two.
It should be noted that the zeroth-order equations coincide with (\ref{se}) and (\ref{tau}) and
their only acceptable solutions are with $p=-\rho$ or without the matter.
Simple manipulations give
\begin{equation}
 F_{x}(\lambda) = \frac{3 H_{0}+\lambda}{3 H_{0}} F_{t}(\lambda)
 \label{FT}
\end{equation}
and for $\sigma \neq -1$ 
\begin{equation}
 F(\lambda) = \left( \sigma + \frac{3 H_{0}+\lambda}{3 H_{0}}\right) F_{t}(\lambda),
 \label{FX}
\end{equation}
where $\sigma$ is the coefficient of the linear equation of state $p=\sigma \rho.$
It should be noted that for $H_{0} >0$ the polynomial $F_{x}(\lambda)$ does not give
more information than $F_{t}(\lambda).$ From Eq. (\ref{FX}) one has $\lambda = -3 H_{0} (1+\sigma),$
which is negative for nontachyonic matter. The problem reduces to the analysis of the roots 
of the equation 
\begin{equation}
F_{t}(\lambda)=\sum_{k=0}^{4} b_{4-k} \lambda^k =0,
\label{eqF}
\end{equation}
where the real coefficients $b_{k}=b_{k}(a_{1},...,a_{10},m,H_{0})$ depend on type 
of the field and $H_{0}.$
It can be shown by a direct calculation  that taking $b_{0}$ to be positive  
the coefficient $b_{3}$ is  always negative for spinor and vector field. The massive
scalar field is more complicated, nevertheless, it is impossible to have all the 
coefficients simultaneously positive. It can be seen that regardless of the exact 
value of $H_{0}$ there is no stable solutions. Consequently, one can draw a conclusion
that there are no stable solutions for the massive spinors, vectors and scalars 
with arbitrary coupling parameter $\xi.$~\footnote{We confirmed this result solving the fourth-order
equation~(\ref{eqF}).}

Although our analyses are devoted solely to the massive fields, it is instructive, 
for a comparison,  to consider quantized massless conformally invariant fields.
This case has been studied in Ref.~\cite{tomislav} to which the reader is referred for details.
The trace of the stress-energy tensor considered in Ref.~\cite{tomislav} has the form~\cite{mazur,mottola}
\begin{equation}
 T_{a}^{a} = b F +b' (E-\frac{2}{3}\Box R)+b'' \Box R,
\end{equation}
where $F =C_{abcd} C^{abcd},$ 
\begin{equation}
 E = R_{abcd} R^{a b c d} -4 R_{ab}R^{ab} +R^{2},
\end{equation}
 is the Gauss-Bonnet invariant  and the numeric
coefficients $b,$ $b'$ and $b''$ depend on the type and number of fields of a given
spin.

Because of the high symmetry of the Friedman-Robertson-Walker spacetime the conservation
equation can be integrated and the stress-energy tensor of the quantized fields 
can be constructed.  Now, setting all time derivatives of $H(t)$ in the semiclassical equations to zero
one obtains two equations:
\begin{equation}
 -3 H^2_{0}-48 \pi b' H_{0}^{4} + \Lambda=-8\pi \rho,
 \label{t0}
\end{equation}
\begin{equation}
 -3 H^2_{0}-48 \pi b' H_{0}^{4} + \Lambda=8\pi p.
 \label{t1}
\end{equation}
Note that the remarks made earlier when discussing the solutions of Eqs. (\ref{se}) and 
(\ref{tau}) are also relevant here. The stability analysis leads to the equations similar 
to (\ref{FT}) and (\ref{FX}) with only one difference: Now the characteristic 
polynomial $F_{t}(\lambda)$ has the form
\begin{equation}
 F_{t}(\lambda) = -8\pi (3b''-2 b') \lambda^{2} -24 \pi H_{0} (3 b''-2 b') \lambda + 3+96 \pi b' H_{0}^{2}.
 \label{fft}
\end{equation}
Upon dividing the characteristic equation by $b_{0}$ one concludes that 
$b''-2/3 b'$ and $1+32 \pi b' H_{0}^2$
should have the opposite signs. A closer analysis carried out for a small cosmological
constant indicates that when $b''-2/3 b'<0$ the classical attractor is unstable whereas 
the quantum attractor is stable. Similarly, when $b''-2/3 b'>0$
the classical attractor is stable and the quantum attractor is unstable. 
The attractors considered here are the solutions of the zeroth-order equations (\ref{t0}) and (\ref{t1}).

The characteristic polynomial (\ref{fft}) contains, as a special case, the results of Ref.~\cite{shapiro3}.
Indeed, putting $\tilde{b} = b'$ and $\tilde{c} =b''-2/3 b'$ and  $\Lambda=0$ after some simplifications
and rearrangements, one obtains
\begin{equation}
 \lambda^{2} + 3 H_{0}^2 \lambda + \frac{1}{8\pi \tilde{c}}=0
 \label{upr}
\end{equation}
and the behavior of system is governed by the sign of $\tilde{c},$ which should be 
positive.\footnote{Note that the authors of
Ref.~\cite{shapiro3} used the value of $\tilde{c}$ as predicted by $\zeta-$function or
point-splitting regularization rather that the value predicted by the dimensional 
regularization.} In simplifying (\ref{fft}) to (\ref{upr}) we used the vacuum solution of (\ref{t0})
\begin{equation}
 H_{0}^{2} = -\frac{1}{16 \pi \tilde{b}}.
\end{equation}

For the massless (conformally invariant) quantum fields one can use 
the trace equation simply because the trace is given by the linear
combinations of the curvature invariants. Although this method
does not work for the massive fields it is possible to use a procedure
called conformization~\cite{shapiro2}. The resulting conformal anomaly
tensor is composed of the Gauss-Bonnet topological invariant,
the square of the Weyl tensor,  $\Box R$ and the auxiliary field $\chi$ introduced
to  make the action invariant under the conformal transformation:
\begin{equation}
 T_{a}^{a} = \tilde{w} F + \tilde{b} E + \tilde{c} \Box R 
 + \tilde{f}\left(R \chi^{2} + 6 (\partial \chi)^2 \right),
\end{equation}
where  $\tilde{w}, \tilde{b}, \tilde{c}$ are the standard coefficients 
(see, e.g., Ref.~\cite{shapiro2}) and   $\tilde{f}$ depends on the number of massive 
spinor fields and their masses. 
Subsequently, after construction of the anomaly-induced effective action one can construct
equations of motion. Now, treating the terms proportional to $\tilde{f}$
as small perturbation one concludes  that fermion anomaly-induced inflation automatically slows down~\cite{shapiro2}.
In our next example we shall show that exponential expansion also slows down 
for $\tau <0.$

Let us consider the semicalssical equations~(\ref{semiE}) 
with $\Lambda=0$ and  the classical matter  characterized by the equation 
of state of the form $p =\sigma \rho.$ The matter density of the classical matter 
evolves according to the formula
\begin{equation}
 \rho(t) = \frac{C_{1}}{a(t)^{3(1+\sigma)}}
\end{equation}
We shall treat the quantum part of the right hand side of the semiclassical 
Einstein field equations as small perturbation. Now, inserting the parameter 
$\varepsilon$ in front of the quantum part of the total stress-energy tensor 
and expanding the scale factor as
\begin{equation}
 a(t) = a_{(0)}(t) + \varepsilon a_{(1)}(t) + {\cal O}(\varepsilon^2)
\end{equation}
one obtains two simple differential equations. The first one 
\begin{equation}
 a'_{(0)} - \omega a_{(0)}^{-\tilde{\sigma}} =0
 \end{equation}
can easily be solved: for $\tilde{\sigma} \neq -1$ one has
\begin{equation}
a_{(0)}(t) = \left[ (\tilde{\sigma} +1) (\omega t+C_{2})\right]^{\frac{1}{\tilde{\sigma}+1}},
\end{equation}
whereas for $\tilde{\sigma} =-1$ one has the exponential law of the evolution of the scale factor
\begin{equation}
 a_{(0)}(t) =C_{3} \exp(\omega t), 
\end{equation}
where $\tilde{\sigma} =\frac{1}{2}(3 \sigma+1),$ $\omega^{2} = 8\pi C_{1}/3,$  and $C_{2}$ and $C_{3}$ are the integration constants.
The second equation is slightly more complicated and can be written in the form: 
\begin{equation}
 a_{(1)}' + \omega \tilde{\sigma} a_{(0)}^{-(1+\tilde{\sigma})} a_{(1)} -\frac{1}{\pi m^{2}} \omega^{5}
 a_{(0)}^{-(4 +5 \tilde{\sigma})}\sum_{n=0}^{4} B_{n}\tilde{\sigma}^{n} =0,
 \label{second}
\end{equation}
where the coefficients $B_{n}$ which depend on the spin of the field are tabulated in Table II.
 \begin{table}
\begin{tabular}{|c|c|c|c|c|c|}\hline 
× & $B_{0}$ & $B_{1}$ & $B_{2}$ & $B_{3}$ & $B_{4}$\\ \hline\hline
$\xi=0$ & 13/144 & -131/420 & -1343/5040 & 2371/7560 & 17/84\\
$\xi=1/6$ & 0 & -2/315 & -1/252 & 41/3780 & 1/126\\
$s =1/2$ & 0 & -17/2520 & -11/10080 & 281/15120 & 1/84\\
$s=1$ & 1/144 & -101/1260 & -151/5040 & 439/2520 & 3/28\\ \hline \hline
                                           \end{tabular}
\caption{The coefficients $B_{n}$ calculated for the massive scalar field with minimal and conformal
curvature coupling, massive spinor and massive vector fields}
 \label{table2}
                                           \end{table}
It can be solved to yield 
\begin{equation}
 a_{(1)} = (\omega t+ C_{2})^{-\frac{\si}{1+\si}} C_{4}-
 (1+\si)^{-\frac{4+5\si}{1+\si}} (\omega t +C_{2})^{-\frac{3+4\si}{1+\si}} \frac{\omega^{4}}{3 \pi m^{2}} 
 \sum_{n=0}^{4} B_{n}\tilde{\sigma}^{n},
\end{equation}
where $C_{4}$ is another integration constant. The $\si =-1$ case has to be treated separately
and after massive simplifications in (\ref{second}) one obtains
\begin{equation}
 a_{1}(t) =C_{5} \exp(\omega t) +\frac{1}{2} C_{3} \omega^{5} \tau t \exp(\omega t),
\end{equation}
where $\tau$ is given by (\ref{tau}).
Further, introducing new constant $\tilde{C}_{3}$  by means of the finite renormalization 
$C_{3} \to \tilde{C}_{3} =C_{3} + \varepsilon C_{5}$ one gets 
\begin{equation}
 a(t) = \tilde{C}_{3} \exp(t\omega)  +\frac{1}{2}\varepsilon \tilde{C}_{3} \omega^{5} \tau t \exp(\omega t),
\end{equation}
where $\tau$ is given as before by (\ref{tau}).
The constant $\tilde{C}_{3}$ may be related to the quantum-corrected ``observed'' scale factor $a(t_{0}).$ 
Simple manipulations give 
\begin{equation}
 a(t) = a(t_{0}) \exp(\omega(t-t_{0}))\left[ 1+\frac{1}{2}\varepsilon \omega^{5} \tau (t-t_{0})\right] + {\mathcal O}(\varepsilon^{2})
\end{equation}
and within the accuracy of our calculations,  ${\mathcal O}(\varepsilon^{2}) ,$ the term in the square brackets is equal
\begin{equation}
 \exp(\frac{1}{2}\varepsilon \omega^{5} \tau (t-t_{0})).
\end{equation}
Inspection of Eq.~(\ref{tau}) shows that the quantum effects tend to decrease  
the rate of expansion for the massive spinors and vectors as well as the massive scalar fields with 
$\xi >\xi_{crit}.$ For the massive scalar field with $\xi <\xi_{crit}$ the rate of the expansion is increased.

There are other families of  (approximate) solutions  which can easily be  constructed 
simply by retaining the cosmological term or allowing the renormalized coupling 
constants to be nonzero or accepting some nonstandard equations of state, but  we shall 
not dwell on them here. We only mention that the procedure adopted in the second example  is
equivalent to the approach of Ref.~\cite{Simon}. This  can 
readily be verified by a direct computation. Indeed, setting the quantum part of the 
stress-energy tensor to zero while retaining the quadratic terms given by (\ref{kwa}) and (\ref{kwb}),
expanding the thus obtained equations and collecting the terms with the like powers of $\varepsilon,$
and finally constructing solutions one gets precisely the results obtained in Ref.~\cite{Simon}.

\section{Final remarks}
In this paper the renormalized stress-energy tensor of the quantized massive scalar, spinor and vector fields
in the spatially flat Friedman-Robertson-Walker universe has been constructed within the framework of 
the Schwinger-DeWitt technique. For the scalar field (with an arbitrary curvature coupling) it reduces 
to the tensor constructed using the adiabatic regularization. There is a clear correspondence between the
both methods: To calculate the first-order approximation of the tensor one needs the 6-order WKB approximation 
of the mode functions in the adiabatic method, or, in the Schwinger-DeWitt method,  the functional derivative of 
the effective action constructed from coincidence limit of the Hadamard-DeWitt coefficient [$a_{3}]$ with respect 
to the metric tensor.
\footnote{We have gone a step further and calculated (for $k=0$ and $ k=\pm 1$) 
the next-to-leading term of the renormalized stress-energy tensor using the effective action constructed 
from the coincidence limit of the coefficient $a_{4}$ on the one hand, and the 8-th order adiabatic 
approximation to the mode function, on the other. Both methods  yield identical results. 
An interesting lesson from this calculations is that the adiabatic method was less time-consuming, 
at least in our implementation of the both algorithms. Since the calculations have been carried out for 
massive scalars only they are somewhat beyond the scope of the present paper and we shall present the 
details of the calculations in a separate publication. The next-to-leading term of the stress-energy 
tensor that has been used in the calculations is given in Refs.~\cite{mat3,mat4}} Such coincidences 
have been found  earlier in the black hole physics~\cite{Samuel}. For the $s=1/2$ and $s=1$  field
we have found an agreement between the stress-energy tensor obtained using the Zeldovich-Starobinsky method 
and Eqs. (\ref{comp_tt}) and (\ref{comp_xx}) of the present paper.
 
With the stress-energy tensor which functionally depends on the scale factor $a(t)$ one can 
attempt to solve the semi-classical Einstein field equations in a self-consistent way. 
We have illustrated the procedure in the two interesting and important cases constructing
particular class of the exact solution of the semi-classical equations and treating the problem
perturbatively. We have preformed stability analysis of our solutions and, for a comparison, we 
briefly discussed earlier results.


Finally, it should be noted, that the adiabatic calculations can  be extended to both $k=-1$ and $k=1$ cases 
(in the $k=1$ case one has summation of the mode functions instead of integration which is an 
obstacle in constructing the final compact expressions). Once again it can be demonstrated that 
the Schwinger-DeWitt and the adiabatic methods yield identical results. We shall present and analyze 
this group of problems in a separate publication. 

We conclude this paper with 
the observation that the Schwinger-DeWitt method, although invented in the mid-1960s,
still goes strong,  ranging its domain of applicability from black hole physics to cosmology, and,
despite its limitations it is still the best general approximation available on the market.

\begin{acknowledgments}
We would like to thank D. Tryniecki and M. Telecka for checking some of our calculations.
\end{acknowledgments}

\begin{thebibliography}{30}
\expandafter\ifx\csname natexlab\endcsname\relax\def\natexlab#1{#1}\fi
\expandafter\ifx\csname bibnamefont\endcsname\relax
  \def\bibnamefont#1{#1}\fi
\expandafter\ifx\csname bibfnamefont\endcsname\relax
  \def\bibfnamefont#1{#1}\fi
\expandafter\ifx\csname citenamefont\endcsname\relax
  \def\citenamefont#1{#1}\fi
\expandafter\ifx\csname url\endcsname\relax
  \def\url#1{\texttt{#1}}\fi
\expandafter\ifx\csname urlprefix\endcsname\relax\def\urlprefix{URL }\fi
\providecommand{\bibinfo}[2]{#2}
\providecommand{\eprint}[2][]{\url{#2}}

\bibitem[{\citenamefont{Kaya and Tarman}(2011)}]{Kaya}
\bibinfo{author}{\bibfnamefont{A.}~\bibnamefont{Kaya}} \bibnamefont{and}
  \bibinfo{author}{\bibfnamefont{M.}~\bibnamefont{Tarman}},
  \bibinfo{journal}{JCAP} \textbf{\bibinfo{volume}{1104}}, \bibinfo{pages}{040}
  (\bibinfo{year}{2011}), \eprint{ArXiv:1104.5562}.

\bibitem[{\citenamefont{Parker and Fulling}(1974)}]{Parker1}
\bibinfo{author}{\bibfnamefont{L.}~\bibnamefont{Parker}} \bibnamefont{and}
  \bibinfo{author}{\bibfnamefont{S.}~\bibnamefont{Fulling}},
  \bibinfo{journal}{Phys.Rev.} \textbf{\bibinfo{volume}{D9}},
  \bibinfo{pages}{341} (\bibinfo{year}{1974}).

\bibitem[{\citenamefont{Fulling et~al.}(1974)\citenamefont{Fulling, Parker, and
  Hu}}]{Parker2}
\bibinfo{author}{\bibfnamefont{S.}~\bibnamefont{Fulling}},
  \bibinfo{author}{\bibfnamefont{L.}~\bibnamefont{Parker}}, \bibnamefont{and}
  \bibinfo{author}{\bibfnamefont{B.}~\bibnamefont{Hu}},
  \bibinfo{journal}{Phys.Rev.} \textbf{\bibinfo{volume}{D10}},
  \bibinfo{pages}{3905} (\bibinfo{year}{1974}).

\bibitem[{\citenamefont{Fulling and Parker}(1974)}]{Parker3}
\bibinfo{author}{\bibfnamefont{S.}~\bibnamefont{Fulling}} \bibnamefont{and}
  \bibinfo{author}{\bibfnamefont{L.}~\bibnamefont{Parker}},
  \bibinfo{journal}{Annals Phys.} \textbf{\bibinfo{volume}{87}},
  \bibinfo{pages}{176} (\bibinfo{year}{1974}).

\bibitem[{\citenamefont{Bunch and Davies}(1978)}]{Bunch1}
\bibinfo{author}{\bibfnamefont{T.}~\bibnamefont{Bunch}} \bibnamefont{and}
  \bibinfo{author}{\bibfnamefont{P.}~\bibnamefont{Davies}},
  \bibinfo{journal}{J.Phys.} \textbf{\bibinfo{volume}{A11}},
  \bibinfo{pages}{1315} (\bibinfo{year}{1978}).

\bibitem[{\citenamefont{Bunch}(1980)}]{Bunch2}
\bibinfo{author}{\bibfnamefont{T.}~\bibnamefont{Bunch}},
  \bibinfo{journal}{J.Phys.} \textbf{\bibinfo{volume}{A13}},
  \bibinfo{pages}{1297} (\bibinfo{year}{1980}).

\bibitem[{\citenamefont{Anderson and Parker}(1987)}]{Anderson}
\bibinfo{author}{\bibfnamefont{P.~R.} \bibnamefont{Anderson}} \bibnamefont{and}
  \bibinfo{author}{\bibfnamefont{L.}~\bibnamefont{Parker}},
  \bibinfo{journal}{Phys.Rev.} \textbf{\bibinfo{volume}{D36}},
  \bibinfo{pages}{2963} (\bibinfo{year}{1987}).

\bibitem[{\citenamefont{Landete et~al.}(2013)\citenamefont{Landete,
  Navarro-Salas, and Torrenti}}]{Landete}
\bibinfo{author}{\bibfnamefont{A.}~\bibnamefont{Landete}},
  \bibinfo{author}{\bibfnamefont{J.}~\bibnamefont{Navarro-Salas}},
  \bibnamefont{and} \bibinfo{author}{\bibfnamefont{F.}~\bibnamefont{Torrenti}}
  (\bibinfo{year}{2013}), \eprint{ArXiv:1305.7374}.

\bibitem[{\citenamefont{Parker and Toms}(2009)}]{ParTom}
\bibinfo{author}{\bibfnamefont{L.}~\bibnamefont{Parker}} \bibnamefont{and}
  \bibinfo{author}{\bibfnamefont{D.}~\bibnamefont{Toms}},
  \emph{\bibinfo{title}{Quantum field theory in curved spacetime: quantized
  fields and gravity}} (\bibinfo{publisher}{Cambridge University Press},
  \bibinfo{year}{2009}).
  
  \bibitem[{\citenamefont{Grib et~al.}(1988)\citenamefont{Grib, Mamayev, and
  Mostepanenko}}]{Grib}
\bibinfo{author}{\bibfnamefont{A.~A.} \bibnamefont{Grib}},
  \bibinfo{author}{\bibfnamefont{S.~G.} \bibnamefont{Mamayev}},
  \bibnamefont{and} \bibinfo{author}{\bibfnamefont{V.~M.}
  \bibnamefont{Mostepanenko}}, \emph{\bibinfo{title}{Vacuum Quantum Effects in
  Strong Fields}} (\bibinfo{publisher}{Energoatomizdat},
  \bibinfo{address}{Moscow}, \bibinfo{year}{1988}), \bibinfo{edition}{(in
  Russian)}.

\bibitem[{\citenamefont{DeWitt}(1975)}]{DeWitt}
\bibinfo{author}{\bibfnamefont{B.~S.} \bibnamefont{DeWitt}},
  \bibinfo{journal}{Phys.Rept.} \textbf{\bibinfo{volume}{19}},
  \bibinfo{pages}{295} (\bibinfo{year}{1975}).

\bibitem[{\citenamefont{Zeldovich and Starobinsky}(1972)}]{Yakov}
\bibinfo{author}{\bibfnamefont{Y.}~\bibnamefont{Zeldovich}} \bibnamefont{and}
  \bibinfo{author}{\bibfnamefont{A.~A.} \bibnamefont{Starobinsky}},
  \bibinfo{journal}{Sov.Phys.JETP} \textbf{\bibinfo{volume}{34}},
  \bibinfo{pages}{1159} (\bibinfo{year}{1972}).

\bibitem[{\citenamefont{Parker and Simon}(1993)}]{Simon}
\bibinfo{author}{\bibfnamefont{L.}~\bibnamefont{Parker}} \bibnamefont{and}
  \bibinfo{author}{\bibfnamefont{J.~Z.} \bibnamefont{Simon}},
  \bibinfo{journal}{Phys.Rev.} \textbf{\bibinfo{volume}{D47}},
  \bibinfo{pages}{1339} (\bibinfo{year}{1993}), \eprint{gr-qc/9211002}.

\bibitem[{\citenamefont{Misner et~al.}(1973)\citenamefont{Misner, Thorne, and
  Wheeler}}]{MTW}
\bibinfo{author}{\bibfnamefont{C.~W.} \bibnamefont{Misner}},
  \bibinfo{author}{\bibfnamefont{K.~S.} \bibnamefont{Thorne}},
  \bibnamefont{and} \bibinfo{author}{\bibfnamefont{J.~A.}
  \bibnamefont{Wheeler}}, \emph{\bibinfo{title}{Gravitation}}
  (\bibinfo{publisher}{WH Freeman}, \bibinfo{address}{San Francisco},
  \bibinfo{year}{1973}).

\bibitem[{\citenamefont{Barvinsky and Vilkovisky}(1985)}]{Barvinsky}
\bibinfo{author}{\bibfnamefont{A.}~\bibnamefont{Barvinsky}} \bibnamefont{and}
  \bibinfo{author}{\bibfnamefont{G.}~\bibnamefont{Vilkovisky}},
  \bibinfo{journal}{Phys.Rept.} \textbf{\bibinfo{volume}{119}},
  \bibinfo{pages}{1} (\bibinfo{year}{1985}).

\bibitem[{\citenamefont{Frolov and Zelnikov}(1984)}]{Frolov}
\bibinfo{author}{\bibfnamefont{V.~P.} \bibnamefont{Frolov}} \bibnamefont{and}
  \bibinfo{author}{\bibfnamefont{A.}~\bibnamefont{Zelnikov}},
  \bibinfo{journal}{Phys.Rev.} \textbf{\bibinfo{volume}{D29}},
  \bibinfo{pages}{1057} (\bibinfo{year}{1984}).

\bibitem[{\citenamefont{Avramidi}(1989)}]{Avramidi}
\bibinfo{author}{\bibfnamefont{I.}~\bibnamefont{Avramidi}},
  \bibinfo{journal}{Theor.Math.Phys.} \textbf{\bibinfo{volume}{79}},
  \bibinfo{pages}{494} (\bibinfo{year}{1989}).

\bibitem[{\citenamefont{Avramidi}(2000)}]{IvanBook}
\bibinfo{author}{\bibfnamefont{I.~G.} \bibnamefont{Avramidi}},
  \emph{\bibinfo{title}{Heat kernel and quantum gravity}}
  (\bibinfo{publisher}{Springer-Verlag}, \bibinfo{address}{Berlin},
  \bibinfo{year}{2000}).

\bibitem[{\citenamefont{Matyjasek et~al.}(2006)\citenamefont{Matyjasek,
  Telecka, and Tryniecki}}]{ja2006}
\bibinfo{author}{\bibfnamefont{J.}~\bibnamefont{Matyjasek}},
  \bibinfo{author}{\bibfnamefont{M.}~\bibnamefont{Telecka}}, \bibnamefont{and}
  \bibinfo{author}{\bibfnamefont{D.}~\bibnamefont{Tryniecki}},
  \bibinfo{journal}{Phys.Rev.} \textbf{\bibinfo{volume}{D73}},
  \bibinfo{pages}{124016} (\bibinfo{year}{2006}), \eprint{hep-th/0606254}.

\bibitem[{\citenamefont{Matyjasek}(2000)}]{mat1}
\bibinfo{author}{\bibfnamefont{J.}~\bibnamefont{Matyjasek}},
  \bibinfo{journal}{Phys.Rev.} \textbf{\bibinfo{volume}{D61}},
  \bibinfo{pages}{124019} (\bibinfo{year}{2000}), \eprint{gr-qc/9912020}.

\bibitem[{\citenamefont{Matyjasek}(2001)}]{mat2}
\bibinfo{author}{\bibfnamefont{J.}~\bibnamefont{Matyjasek}},
  \bibinfo{journal}{Phys.Rev.} \textbf{\bibinfo{volume}{D63}},
  \bibinfo{pages}{084004} (\bibinfo{year}{2001}), \eprint{gr-qc/0010097}.

\bibitem[{\citenamefont{Lu and Wise}(1993)}]{Lu}
\bibinfo{author}{\bibfnamefont{M.}~\bibnamefont{Lu}} \bibnamefont{and}
  \bibinfo{author}{\bibfnamefont{M.~B.} \bibnamefont{Wise}},
  \bibinfo{journal}{Phys.Rev.} \textbf{\bibinfo{volume}{D47}},
  \bibinfo{pages}{3095} (\bibinfo{year}{1993}), \eprint{gr-qc/9301021}.

\bibitem[{\citenamefont{Goroff and Sagnotti}(1986)}]{Goroff}
\bibinfo{author}{\bibfnamefont{M.~H.} \bibnamefont{Goroff}} \bibnamefont{and}
  \bibinfo{author}{\bibfnamefont{A.}~\bibnamefont{Sagnotti}},
  \bibinfo{journal}{Nucl.Phys.} \textbf{\bibinfo{volume}{B266}},
  \bibinfo{pages}{709} (\bibinfo{year}{1986}).

\bibitem[{\citenamefont{Dobado and Maroto}(1993)}]{Dobado}
\bibinfo{author}{\bibfnamefont{A.}~\bibnamefont{Dobado}} \bibnamefont{and}
  \bibinfo{author}{\bibfnamefont{A.~L.} \bibnamefont{Maroto}},
  \bibinfo{journal}{Phys.Lett.} \textbf{\bibinfo{volume}{B316}},
  \bibinfo{pages}{250} (\bibinfo{year}{1993}), \eprint{hep-ph/9309221}.

\bibitem[{\citenamefont{Beilin et~al.}(1980)\citenamefont{Beilin, Vereshkov,
  Grishkan, Ivanov, Nesterenko et~al.}}]{Beilin}
\bibinfo{author}{\bibfnamefont{V.}~\bibnamefont{Beilin}},
  \bibinfo{author}{\bibfnamefont{G.}~\bibnamefont{Vereshkov}},
  \bibinfo{author}{\bibfnamefont{Y.}~\bibnamefont{Grishkan}},
  \bibinfo{author}{\bibfnamefont{N.}~\bibnamefont{Ivanov}},
  \bibinfo{author}{\bibfnamefont{V.}~\bibnamefont{Nesterenko}},
  \bibnamefont{et~al.}, \bibinfo{journal}{Sov.Phys.JETP}
  \textbf{\bibinfo{volume}{51}}, \bibinfo{pages}{1045} (\bibinfo{year}{1980}).

\bibitem[{\citenamefont{Xu}(1987)}]{Xu1}
\bibinfo{author}{\bibfnamefont{D.-Y.} \bibnamefont{Xu}},
  \bibinfo{journal}{Phys.Rev.} \textbf{\bibinfo{volume}{D35}},
  \bibinfo{pages}{769} (\bibinfo{year}{1987}).

\bibitem[{\citenamefont{Harvey}(1995)}]{Harvey}
\bibinfo{author}{\bibfnamefont{A.}~\bibnamefont{Harvey}}, \bibinfo{journal}{J.
  Math. Phys.} \textbf{\bibinfo{volume}{36}}, \bibinfo{pages}{356}
  (\bibinfo{year}{1995}).

\bibitem[{\citenamefont{Maroto and Shapiro}(1997)}]{shapiro1}
\bibinfo{author}{\bibfnamefont{A.~L.} \bibnamefont{Maroto}} \bibnamefont{and}
  \bibinfo{author}{\bibfnamefont{I.}~\bibnamefont{Shapiro}},
  \bibinfo{journal}{Phys.Lett.} \textbf{\bibinfo{volume}{B414}},
  \bibinfo{pages}{34} (\bibinfo{year}{1997}), \eprint{hep-th/9706179}.

\bibitem[{\citenamefont{Shapiro and Sola}(2002)}]{shapiro2}
\bibinfo{author}{\bibfnamefont{I.~L.} \bibnamefont{Shapiro}} \bibnamefont{and}
  \bibinfo{author}{\bibfnamefont{J.}~\bibnamefont{Sola}},
  \bibinfo{journal}{Phys.Lett.} \textbf{\bibinfo{volume}{B530}},
  \bibinfo{pages}{10} (\bibinfo{year}{2002}), \eprint{hep-ph/0104182}.

\bibitem[{\citenamefont{Cesari}(1963)}]{cesari}
\bibinfo{author}{\bibfnamefont{L.}~\bibnamefont{Cesari}},
  \emph{\bibinfo{title}{Asymptotic behavior and stability problems in ordinary
  differential equations}}, (\bibinfo{publisher}{Academic Press Inc.}
  \bibinfo{address}{Publishers, New York}, \bibinfo{year}{1963}).

\bibitem[{\citenamefont{Koksma and Prokopec}(2008)}]{tomislav}
\bibinfo{author}{\bibfnamefont{J.~F.} \bibnamefont{Koksma}} \bibnamefont{and}
  \bibinfo{author}{\bibfnamefont{T.}~\bibnamefont{Prokopec}},
  \bibinfo{journal}{Phys.Rev.} \textbf{\bibinfo{volume}{D78}},
  \bibinfo{pages}{023508} (\bibinfo{year}{2008}), \eprint{ArXiv:0803.4000}.

\bibitem[{\citenamefont{Mazur and Mottola}(2001)}]{mazur}
\bibinfo{author}{\bibfnamefont{P.~O.} \bibnamefont{Mazur}} \bibnamefont{and}
  \bibinfo{author}{\bibfnamefont{E.}~\bibnamefont{Mottola}},
  \bibinfo{journal}{Phys.Rev.} \textbf{\bibinfo{volume}{D64}},
  \bibinfo{pages}{104022} (\bibinfo{year}{2001}), \eprint{hep-th/0106151}.

\bibitem[{\citenamefont{Antoniadis et~al.}(2007)\citenamefont{Antoniadis,
  Mazur, and Mottola}}]{mottola}
\bibinfo{author}{\bibfnamefont{I.}~\bibnamefont{Antoniadis}},
  \bibinfo{author}{\bibfnamefont{P.~O.} \bibnamefont{Mazur}}, \bibnamefont{and}
  \bibinfo{author}{\bibfnamefont{E.}~\bibnamefont{Mottola}},
  \bibinfo{journal}{New J.Phys.} \textbf{\bibinfo{volume}{9}},
  \bibinfo{pages}{11} (\bibinfo{year}{2007}), \eprint{gr-qc/0612068}.

\bibitem[{\citenamefont{Pelinson et~al.}(2003)\citenamefont{Pelinson, Shapiro,
  and Takakura}}]{shapiro3}
\bibinfo{author}{\bibfnamefont{A.}~\bibnamefont{Pelinson}},
  \bibinfo{author}{\bibfnamefont{I.}~\bibnamefont{Shapiro}}, \bibnamefont{and}
  \bibinfo{author}{\bibfnamefont{F.}~\bibnamefont{Takakura}},
  \bibinfo{journal}{Nucl.Phys.} \textbf{\bibinfo{volume}{B648}},
  \bibinfo{pages}{417} (\bibinfo{year}{2003}), \eprint{hep-ph/0208184}.



\bibitem[{\citenamefont{Matyjasek and Tryniecki}(2009)}]{mat3}
\bibinfo{author}{\bibfnamefont{J.}~\bibnamefont{Matyjasek}} \bibnamefont{and}
  \bibinfo{author}{\bibfnamefont{D.}~\bibnamefont{Tryniecki}},
  \bibinfo{journal}{Phys.Rev.} \textbf{\bibinfo{volume}{D79}},
  \bibinfo{pages}{084017} (\bibinfo{year}{2009}), \eprint{ArXiv:0901.2746}.

\bibitem[{\citenamefont{Matyjasek et~al.}(2010)\citenamefont{Matyjasek,
  Tryniecki, and Zwierzchowska}}]{mat4}
\bibinfo{author}{\bibfnamefont{J.}~\bibnamefont{Matyjasek}},
  \bibinfo{author}{\bibfnamefont{D.}~\bibnamefont{Tryniecki}},
  \bibnamefont{and}
  \bibinfo{author}{\bibfnamefont{K.}~\bibnamefont{Zwierzchowska}},
  \bibinfo{journal}{Phys.Rev.} \textbf{\bibinfo{volume}{D81}},
  \bibinfo{pages}{124047} (\bibinfo{year}{2010}), \eprint{ArXiv:1005.1427}.

\bibitem[{\citenamefont{Anderson et~al.}(1995)\citenamefont{Anderson, Hiscock,
  and Samuel}}]{Samuel}
\bibinfo{author}{\bibfnamefont{P.~R.} \bibnamefont{Anderson}},
  \bibinfo{author}{\bibfnamefont{W.~A.} \bibnamefont{Hiscock}},
  \bibnamefont{and} \bibinfo{author}{\bibfnamefont{D.~A.}
  \bibnamefont{Samuel}}, \bibinfo{journal}{Phys.Rev.}
  \textbf{\bibinfo{volume}{D51}}, \bibinfo{pages}{4337} (\bibinfo{year}{1995}).

\end{thebibliography}

\end{document}